\documentclass[11pt, a_4paper]{openjournal}

\pdfoutput=1

\usepackage{amsmath}
\usepackage{graphicx}

\usepackage[pdftex, colorlinks=true, linkcolor=black, urlcolor=black, citecolor=blue]{hyperref}

\usepackage[english]{babel}
\usepackage[utf8]{inputenc}
\usepackage{enumerate}
\usepackage{amsbsy}
\usepackage{amssymb} %used by default

\usepackage{amsfonts}
\usepackage{pstricks}
\usepackage{color}
\usepackage{setspace}

\newcommand{\bea}{\begin{eqnarray}} \newcommand{\eea}{\end{eqnarray}}
\newcommand{\el}{\nonumber \\}
\newcommand{\re}[1]{(\ref{#1})}

\newcommand{\pat}{\partial}

\renewcommand{\sec}[1]{section \ref{#1}}

\newcommand{\tab}[1]{table \ref{#1}}

\renewcommand{\a}{\alpha}
\renewcommand{\b}{\beta}
\renewcommand{\c}{\gamma}

\newcommand{\ha}{\frac{1}{2}}

\newcommand{\ie}{i.e.\ }

\newcommand{\Mpl}{M_{{}_{\mathrm{Pl}}}}
\newcommand{\rmd}{\text{d}}

\begin{document}

\begin{flushleft}
	\hfill		 HIP-2022-31/TH \\
\end{flushleft}

\title{Palatini formulation for gauge theory: implications for slow-roll inflation}

\author{Syksy R\"{a}s\"{a}nen}
\email{syksy.rasanen@iki.fi}
\address{University of Helsinki, Department of Physics and Helsinki Institute of Physics,\\ P.O. Box 64, FIN-00014 University of Helsinki, Finland}

\author{Yosef Verbin}
\email{verbin@openu.ac.il}
\address{Astrophysics Research Center, the Open University of Israel, Raanana 4353701, Israel}

\begin{abstract}

\noindent We consider a formulation of gauge field theory where the gauge field $A_\a$ and the field strength $F_{\a\b}$ are independent variables, as in the Palatini formulation of gravity. For the simplest gauge field action, this is known to be equivalent to the usual formulation. We add non-minimal couplings between $F_{\a\b}$ and a scalar field, solve for $F_{\a\b}$ and insert it back into the action. This leads to modified gauge field and scalar field terms. We consider slow-roll inflation and show that because of the modifications to the scalar sector, adding higher order terms to the inflaton potential does not spoil its flatness, unlike in the usual case. Instead it makes the effective potential closer to quadratic.

\end{abstract}

\maketitle

\setcounter{tocdepth}{2}

\setcounter{secnumdepth}{3}

\section{Introduction} \label{sec:intro}

There are different formulations of general relativity. One of the most studied alternatives to the metric formulation is the Palatini formulation (also called the metric-affine formulation) where the metric and the affine connection are independent variables \cite{einstein1925, ferraris1982}. For the simplest gravitational action (\ie the Einstein--Hilbert action) and minimally coupled matter (\ie matter that does not directly couple to the connection), the equation of motion of the connection gives back the Levi--Civita connection (up to a projective transformation, which is a symmetry of the action). For a more complicated action, the connection is not Levi--Civita, and the two formulations are physically distinct theories.

The Palatini formulation is also known as the first-order formulation, because only first derivatives of the fields appear in the Riemann tensor and hence in Lagrangians constructed algebraically from it, unlike in the metric formulation. Inserting the solution of the connection equation of motion back into the action allows to eliminate the connection and obtain a theory where the metric is the only degree of freedom. The connection is thus sometimes called an auxiliary field, although it is the metric that appears algebraically in the original action, not the connection.

Analogously, electromagnetism and other gauge theories are usually formulated in what we call the gauge field formulation, where the gauge field is the only variable. However, they can also be formulated in the Palatini style by taking the field strength to be an independent variable in addition to the gauge field. In the gauge field formulation, the simplest action is quadratic in the exterior derivative of the gauge field (plus the commutator for non-Abelian theories). In the Palatini formulation, it is split into a term that is quadratic in the field strength and another that is bilinear in the field strength and the gauge field. This formulation is used in the Faddeev--Jackiw formalism to make quantisation easier \cite{Faddeev:1969su, Faddeev:1988qp}. Solving the classical equation of motion and inserting the solution into the action gives back the quadratic action for the gauge field. This procedure is nothing but a Legendre transformation, similar to the one used to turn actions that are non-linear in the Riemann tensor into the Einstein--Hilbert form in the Palatini formulation for gravity \cite{Magnano:1987zz, Koga:1998un, Afonso:2017}.

However, as in the case of gravity, the Palatini formulation is not equivalent to the usual formulation if the gauge field action is more complicated or if other fields couple directly to the field strength. We investigate this possibility for an Abelian gauge field. We keep the action linear in the gauge field, but include linear and quadratic terms in the field strength, with non-minimal couplings to a scalar field. We solve for the field strength to first order in the kinetic term of the scalar field and insert the solution back into the action. We discuss consequences for slow-roll inflation.

In \sec{sec:calc} we present the formulation, write down the action, solve for the field strength and insert it back into the action. In \sec{sec:pol} we consider the case of polynomial non-minimal couplings, in particular Higgs inflation. In \sec{sec:disc} we discuss what would change for a non-Abelian gauge theory, or if we include fermions or higher order terms in the field strength. In \sec{sec:conc} we summarise our findings and mention open issues.

\newpage

\section{Palatini formulation in the gauge sector} \label{sec:calc}

\subsection{Comparison of the Palatini formulation for gravity and gauge fields}

Let us start from the simplest action for an Abelian gauge field $A_\a$,
\bea \label{actionEM}
  S &=& \int\rmd^4 x \sqrt{-g} \left( - A_{\a\b} A^{\a\b} \right) \el
  &=& \int\rmd^4 x \sqrt{-g} \left( - F_{\a\b} A^{\a\b} + \frac{1}{4} F_{\a\b} F^{\a\b} \right) \ ,
\eea
where because of invariance under the gauge transformations $A_\a\to A_\a+\pat_\a\sigma$ ($\sigma$ is an arbitrary function) the gauge field appears only in the combination $A_{\a\b}\equiv\pat_{[\a} A_{\b]}$. On the second line we have introduced the field strength $F_{\a\b}$. If we consider the Palatini formulation, where the field strength is an independent variable, then varying the action with respect to $F_{\a\b}$ gives the equation of motion $F_{\a\b}=2A_{\a\b}$. Inserting this back into the action restores the form on the first line. Actions of the form on the second line that are linear in the derivatives of $A_\a$ may be more convenient for quantisation, as in the Faddeev--Jackiw formalism \cite{Faddeev:1969su, Faddeev:1988qp}.

Analogously, we can consider gravity in the Palatini formulation, where the connection is an independent variable. For the simplest action, the Einstein--Hilbert action, we have (we use units where the reduced Planck mass is unity)
\bea \label{actiongravity}
  S &=& \int\rmd^4 x \sqrt{-g} \ha g^{\a\b} R_{\a\b} \el
  &=& \int\rmd^4 x \sqrt{-g} g^{\a\b} \left( \pat_{[\mu} \Gamma^\mu_{\b]\a} + \Gamma^\nu{}_{[\nu|\mu|}\Gamma^\mu{}_{\b]\a} \right) \el
  &=& \int\rmd^4 x \sqrt{-g} g^{\a\b} \left( \ha \mathring R_{\a\b} + \mathring \nabla_{[\mu} L^\mu{}_{\b]\a} + L^\nu{}_{[\nu|\mu|} L^\mu{}_{\b]\a} \right) \ ,
\eea
where on the second line we have expressed the Ricci tensor $R_{\a\b}$ in terms of the connection $\Gamma^\c_{\a\b}$, and on the third line we have decomposed the connection as $\Gamma^\c_{\a\b}=\mathring\Gamma^\c_{\a\b}+L^\c{}_{\a\b}$, where $\mathring\Gamma^\c_{\a\b}$ is the Levi--Civita connection and $L^\c{}_{\a\b}$ is the disformation tensor; $\mathring{}$ refers to quantities defined with the Levi--Civita connection. As derivatives of $L^\c{}_{\a\b}$ appear only in the linear term, which is a total derivative, the equation of motion for $L^\c{}_{\a\b}$ is algebraic. Thus the connection does not involve any new degrees of freedom. For the Einstein--Hilbert action the equation of motion shows that $L^\c{}_{\a\b}$ vanishes up to the projective transformation $L^\c{}_{\a\b}\to L^\c{}_{\a\b}+\delta^\c{}_\b\sigma_\a$ ($\sigma_\a$ is an arbitrary vector). The projective transformation does not change the action and is unphysical.\footnote{We could analogously write $F_{\a\b}=2 A_{\a\b}+C_{\a\b}$ on the second line of \re{actionEM}, giving the Lagrangian $-A_{\a\b} A^{\a\b}+\frac{1}{4}C_{\a\b} C^{\a\b}$. It is then transparent that variation with respect to $C_{\a\b}$ gives $C_{\a\b}=0$, and there is no counterpart to the accidental projective symmetry of the gravity sector.} We could equivalently vary with respect to the full $\Gamma^\c_{\a\b}$, obtain the Levi--Civita connection (up to the projective transformation) and insert it back into the action to recover the Einstein--Hilbert action in the metric formulation.

A gravity action that is non-linear in the derivatives of the connection can lead to new dynamical degrees of freedom. Alternatively, it can simply change the relation between existing degrees of freedom, as in the case when the action depends non-linearly on the Ricci scalar \cite{ShahidSaless:1987, Sotiriou:2006, Sotiriou:2008, Olmo:2011}. If we include matter coupled to the connection (but not its derivatives), then the algebraic equation of motion in general gives a non-zero solution for $L^\c{}_{\a\b}$, which, when inserted back into the action, leads to modified interactions for matter.

The dependence on the dynamical field (the metric or the gauge field) is more complicated in the gravity case than in the gauge case, but the essential similarity is that both actions \re{actionEM} and \re{actiongravity} are quadratic in the field strength. If we include terms that couple other fields directly to $F_{\a\b}$, the solution will no longer be $F_{\a\b}=2 A_{\a\b}$, and the equation of motion for $A_{\a\b}$ will be modified, in analogy to the gravity sector in the Palatini case. As $F_{\a\b}$ is an independent field, it does not have to be antisymmetric like $A_{\a\b}$. Analogously, in the gravitational Palatini formulation it is not necessary for the connection to be symmetric as in the metric case.\footnote{There are also versions of the Palatini formulation where constraints are imposed on the connection a priori. The most common constraints are taking the connection to be symmetric, or the non-metricity to be zero. The former condition is sometimes considered to be part of the definition of the Palatini formulation, with the term metric-affine formulation reserved for the case without constraints. A theory with the latter condition is known as Einstein--Cartan gravity. Another common constraint is taking the non-metricity tensor to be proportional to a vector field, $\nabla_\c g_{\a\b}=V_\c g_{\a\b}$.} We compare the gravity and the gauge case in \tab{tab:comp}, starting with the symmetry of the dynamical field that constrains how it may appear in the action. Let us now look at how non-minimal couplings affect the symmetric and antisymmetric parts of $F_{\a\b}$.

\vspace{-1.5cm}

\begin{table*}[t!]
\begin{center}
\begin{tabular}{|c|c|c|c|}
\hline
& gravity sector & gauge sector \\
\hline
dynamical field & $g_{\a\b}$ & $A_\a$ \\
assumed symmetry & $x^\a\to x'{}^\a(x)$ & $A_\a\to A_\a+\pat_\a\sigma$ \\
accidental symmetry & $\Gamma^\c_{\a\b}\to\Gamma^\c_{\a\b}+\delta^\c{}_\b \sigma_\a$ & $-$ \\
Palatini field strength & $\Gamma^\c_{\a\b}$ & $F_{\a\b}$ \\
usual field strength & $\mathring\Gamma^\c_{\a\b}$ & $2 \pat_{[\a} A_{\b]}$ \\
change in the usual components of the field strength & $\Gamma^\c_{(\a\b)}-\mathring\Gamma^\c_{\a\b}$ & $F_{[\a\b]}-2 \pat_{[\a} A_{\b]}$ \\
new field strength components & $\Gamma^\c_{[\a\b]}$ & $F_{(\a\b)}$ \\
\hline
\end{tabular}
\end{center}
\caption{Comparison of the Palatini formulation in the gravity and in the gauge sector.}
\label{tab:comp}
\end{table*}
\vspace{2cm}

\subsection{Non-minimal coupling to a scalar field}

We generalise the action \re{actionEM} by including direct couplings between the field strength and a real scalar field. In the gravity sector, we include a non-minimal coupling to the Ricci scalar. We assume that gravity too is described by the Palatini formulation; in \sec{sec:pol} we comment on what would change in the metric formulation. We will later take the gauge field and the scalar field to be the $U(1)$ field and the Higgs field of the Standard Model, respectively, and consider implications for Higgs inflation \cite{Bezrukov:2007, Bauer:2008} (for reviews, see \cite{Bezrukov:2013, Bezrukov:2015, Rubio:2018ogq}). For the moment they remain general. In the gauge sector, we keep the action linear in $A_\a$, and consider all algebraic terms up to second order in the field strength $F_{\a\b}$ and $A_\a$, and at most linear in the kinetic term of the scalar field. The action is
\bea \label{actionJ}
  S &=& \int\rmd^4 x \sqrt{-g} \Big\{ \ha f(\varphi) R - \ha X - V(\varphi) + a_1(\varphi, X) A_{\a\b} F^{\a\b} + a_2(\varphi, X) {}^*A_{\a\b} F^{\a\b} \el
  && + \left[ a_3(\varphi) A^\a{}_\mu F^{\mu\b} + a_4(\varphi) A^\a{}_\mu F^{\b\mu} + a_5(\varphi) {}^*A^\a{}_\mu F^{\mu\b} + a_6(\varphi) {}^*A^\a{}_\mu F^{\b\mu} \right] X_{\a\b} \el
  && + \tilde b_1(\varphi, X) F^{\a\b} F_{\a\b} + \tilde b_2(\varphi, X) F^{\a\b} F_{\b\a} + b_3(\varphi, X) ^* F^{\a\b} F_{\a\b} + c_1(\varphi, X) F^\a{}_\a + c_2(\varphi, X) (F^\a{}_\a)^2 \el
  && + \left[ d_1(\varphi) F^{\a\b} + d_2(\varphi) F^\mu{}_\mu F^{\a\b} + d_3(\varphi) F^{\a\mu} F_\mu{}^\b + d_4(\varphi) F^{\mu\a} F_\mu{}^\b + d_5(\varphi) F^{\a\mu} F^\b{}_\mu \right. \el
  && \left. + d_6(\varphi) ^*F^{\a\mu} F_\mu{}^\b + d_7(\varphi) ^*F^{\a\mu} F^\b{}_\mu \right] X_{\a\b} \Big\} \ ,
\eea
where $R\equiv g^{\a\b} R_{\a\b}$, $X_{\a\b}\equiv\pat_\a \varphi \pat_\b \varphi$, $X\equiv g^{\a\b} X_{\a\b}$, and $^*F^{\a\b}\equiv\ha\epsilon^{\a\b\mu\nu}F_{\mu\nu}$. By our assumptions, the coefficient functions are at most linear in $X$. While $\varphi$ and $A_{\a\b}$ are even and $^*A_{\a\b}$ is odd under parity, the transformation of $F_{\a\b}$ under parity is not defined a priori, it is determined by the solution to the equation of motion. We decompose the first two terms on the second line of \re{actionJ} as $\tilde b_1 F^{\a\b} F_{\a\b} + \tilde b_2 F^{\a\b} F_{\b\a} = b_1 F^{(\a\b)} F_{(\a\b)} + b_2 F^{[\a\b]} F_{[\a\b]}$, where $b_1\equiv \tilde b_1+\tilde b_2$ and $b_2\equiv \tilde b_1-\tilde b_2$. This makes it transparent that only the $d_i$ terms (with $i=3\ldots7$) couple the symmetric part $F_{(\a\b)}$ and the antisymmetric part $F_{[\a\b]}$. The terms linear in $F_{\a\b}$, with coefficients $a_i$, $c_1$ and $d_1$, give the source terms for $F_{\a\b}$.

Minimising the action with respect to $F_{\a\b}$ gives the equations
\bea \label{eom}
  0 &=& 2 b_1 F_{(\a\b)} + ( c_1 + 2 c_2  F ) g_{\a\b} + ( d_1 + d_2 F ) X_{\a\b} + d_2 F^{\mu\nu} X_{\mu\nu} g_{\a\b} + ( d_3 + d_4 + d_5 ) ( F_{(\a\mu)} X^\mu{}_\b + F_{(\b\mu)} X^\mu{}_\a ) \el
  && + ( d_4 - d_5 ) ( F_{[\a\mu]} X^\mu{}_\b + F_{[\b\mu]} X^\mu{}_\a ) + ( d_6 + d_7 ) ^*F_{\mu(\a} X^\mu{}_{\b)} + ( a_3 + a_4 ) A_{\mu(\a} X^\mu{}_{\b)} + ( a_5 + a_6 ) {}^*A_{\mu(\a} X^\mu{}_{\b)} \\
  0 &=& a_1 A_{\a\b} + a_2 {}^*A_{\a\b} + 2 b_2 F_{[\a\b]} + 2 b_3 {}^*F_{\a\b} + ( a_3 - a_4 ) A_{\mu[\a} X^\mu{}_{\b]} + ( a_5 - a_6 ) {}^*A_{\mu[\a} X^\mu{}_{\b]} \el
  && + ( d_4 - d_5 ) ( F_{(\a\mu)} X^\mu{}_\b - F_{(\b\mu)} X^\mu{}_\a ) + ( - d_3 + d_4 + d_5 ) ( F_{[\a\mu]} X^\mu{}_\b - F_{[\b\mu]} X^\mu{}_\a ) + ( d_6 - d_7 ) ^*F^\mu{}_{[\a} X_{\b]\mu} \el
  && - \ha \epsilon_{\a\b}{}^{\mu\nu} [ ( d_6 + d_7 ) F_{(\mu\rho)} + ( d_6 - d_7 ) F_{[\mu\rho]} ] X^\rho{}_\nu \ ,
\eea
where $^*A_{\a\b}\equiv\ha\epsilon_{\a\b\mu\nu}A^{\mu\nu}$. The general solution of these linear equations is rather involved. We only consider the solution to first order $X_{\a\b}$, which is sufficient for slow-roll inflation.\footnote{Kinetic terms beyond linear order in $X_{\a\b}$ could be useful for kinetically driven inflation, and it would be straightforward to include higher order terms in the action and the solution \cite{Armendariz-Picon:1999hyi, Garriga:1999vw}.} We can then solve the equations iteratively, with the result
\bea \label{Fsol}
  F_{(\a\b)} &=& - \frac{c_1}{2 b_1 + 8 c_2} g_{\a\b} + \frac{ 2 ( b_1 + 4 c_2 ) c_2 d_1 + b_1 c_1 d_2 - 2 c_1 c_2 ( 2 d_2 + d_3 + d_4 + d_5 ) }{4 b _1 (  b_1 + 4 c_2 )^2} X g_{\a\b} \el
  && - \frac{ ( b_1 + 4 c_2 ) d_1 - c_1 ( 2 d_2 + d_3 + d_4 + d_5 ) }{ 2 b_1 ( b_1 + 4 c_2 ) } X_{\a\b} \el
  && + \frac{ 2 \a_1 (d_4 - d_5) - \a_2 (d_6 + d_7) - a_3 - a_4 }{2 b_1} A_{\mu(\a} X^\mu{}_{\b)} + \frac{ 2 \a_2 (d_4 - d_5) - \a_1 (d_6 + d_7) - a_5 - a_6 }{2 b_1} {}^*A_{\mu(\a} X^\mu{}_{\b)} \el
  F_{[\a\b]} &=& \a_1 A_{\a\b} + \a_2 {}^*A_{\a\b} + ( \a_3 A_{\mu[\a} + \a_4 {}^*A_{\mu[\a} ) X^\mu{}_{\b]} + \ha \a_5 \epsilon_{\a\b}{}^{\mu\nu} A_{\mu\rho} X^\rho{}_\nu \ .
\eea
The coefficients $\a_i$ are lengthy functions of $a_i$, $b_i$, and $d_i$, and we do not need them for our slow-roll inflation analysis, so they are relegated to appendix \ref{app:ab}. Note that $\a_1$ and $\a_2$ include terms proportional to $X$. The field strength $F_{\a\b}$ has a mixture of even and odd parity terms. For minimal coupling, we would have $\a_i=0$ for $i\geq3$, $F_{(\a\b)}=0$, and only $A_{\a\b}$ and its dual $^*A_{\a\b}$ would remain in the solution for $F_{[\a\b]}$, with constant coefficients. Were we to impose the constraint that $F_{\a\b}$ is antisymmetric, we would get only the second equation in \re{Fsol}. (To linear order in $X_{\a\b}$, the solution for the antisymmetric part does not depend on the couplings of the symmetric part; the reverse is not true. This does not hold at second order and beyond.) This would be analogous to the gravitational Palatini formulation in the case where the connection is taken to be symmetric. There the non-metric contributions modify the symmetric part of the connection, and in the gauge case, the antisymmetric part of the field strength would also deviate from the standard result $F_{\a\b}=2 A_{\a\b}$ due to the non-minimal couplings.

We do not impose any constraints on $F_{\a\b}$, and insert \re{Fsol} back into the action \re{actionJ}, which then involves only $\varphi$ and $A_\a$. We expand $b_i(\varphi,X)=b_i^{(0)}(\varphi)+b_i^{(1)}(\varphi)X$, and correspondingly for $c_i$. The action is then, to linear order in $X_{\a\b}$,
\bea \label{actionE}
  S &=& \int\rmd^4 x \sqrt{-g} \left[ \ha f R - \ha X - \ha \Delta K X - V - \Delta V + \b_1 A_{\a\b} A^{\a\b} + \b_2 A_{\a\b} {}^*A^{\a\b} + \left( \b_3 A^{\a\mu} A^\b{}_\mu + \b_4 A^{\a\mu} {}^*A^\b{}_\mu \right) X_{\a\b} \right] \el
  &=& \int\rmd^4 x \sqrt{-g} \left[ \ha R - \ha \frac{1 + \Delta K}{f} X - \frac{V + \Delta V}{f^2} + \b_1 A_{\a\b} A^{\a\b} + \b_2 A_{\a\b} {}^*A^{\a\b} + \left( \b_3 A^{\a\mu} A^\b{}_\mu + \b_4 A^{\a\mu} {}^*A^\b{}_\mu \right) X_{\a\b} \right] \el
  &\equiv&  \int\rmd^4 x \sqrt{-g} \left[ \ha R - \ha K X - U + \b_1 A_{\a\b} A^{\a\b} + \b_2 A_{\a\b} {}^*A^{\a\b} + \left( \b_3 A^{\a\mu} A^\b{}_\mu + \b_4 A^{\a\mu} {}^*A^\b{}_\mu \right) X_{\a\b} \right] \ ,
\eea
where the gauge field term coefficients $\b_i$ (which we do not need for analysis of slow-roll inflation) are given in appendix \ref{app:ab}. In the second equality we have carried out the conformal transformation $g_{ab}\to f^{-1} g_{\a\b}$ to make the scalar field minimally coupled.\footnote{Note that we cannot shift the direct coupling of the scalar field to $A_{\a\b} A^{\a\b}$ to the scalar sector as we do with the direct coupling to $R$, as $\sqrt{-g} A_{\a\b} A^{\a\b}$ is invariant under the conformal transformation.} In the last equality we have defined $K\equiv(1+\Delta K)/f$, $U\equiv(V+\Delta V)/f^2$. This action contains both even and odd parity terms. The additive contributions to the kinetic term and the potential coming from the non-minimal coupling to the field strength are
\bea \label{KandU}
  \Delta K &=& - c_1^{(0)} \frac{ - 2 ( b_1^{(0)} + 4 c_2^{(0)} ) d_1 + c_1^{(0)} ( 4 d_2 + d_3 + d_4 + d_5 ) + 4 c_1^{(0)} ( b_1^{(1)} + 4 c_2^{(1)} ) - 8 c_1^{(1)} ( b_1^{(0)} +4 c_2^{(0)} ) }{ 2 (b_1^{(0)} + 4 c_2^{(0)})^2 } \el
  \Delta V &=& \frac{ c_1^{(0)}{}^2}{ b_1^{(0)} + 4 c_2^{(0)} } \ .
\eea
For stability, the coupling functions have to be such that $U$ is bounded from below. We should also have $K>0$, unless we include higher order terms in $X_{\a\b}$, which could make the sum of the kinetic sector terms positive even if $K<0$.

If the scalar field is constant, then only the terms $A_{\a\b} A^{\a\b}$ and $A_{\a\b} {}^*A^{\a\b}$ (and vacuum energy) remain, with constant coefficients. The latter is a total derivative, and so does not affect the equations of motion. The former gives the standard $U(1)$ Lagrangian (up to an irrelevant constant rescaling of $A_{\a\b}$). If the scalar field is not constant, the non-minimal couplings break the conformal symmetry of the gauge field, and could be used for magnetogenesis, with the parity odd terms leading to helical magnetic fields \cite{Durrer:2013pga}. They could also be important during reheating, and when calculating loop corrections.

Let us compare the effect of the non-minimal couplings to the field strength in the gravity and in the gauge sector on the scalar field. In the Palatini formulation for gravity, the non-minimal coupling $f$ to the Ricci scalar is shifted to the scalar sector where it manifests as the multiplicative factors $1/f$ in the kinetic term and $1/f^2$ in the potential. In contrast, the non-minimal couplings to the gauge field strength manifest as an additive term in both the kinetic term and the potential. Modifications in either sector lead to novel effects on their own, and also have interesting interplay. Let us consider this in the case of polynomial coupling functions.

%\vspace{0.5cm}

\section{The case of polynomial coupling functions} \label{sec:pol}

We consider inflation at the classical level, and put $A_\a=0$. The scalar field kinetic term in \re{KandU} contains 11 coupling functions, and the potential depends on 3 functions. The behaviour of the theory depends strongly on how these functions are chosen. We look at the case when all the functions are polynomial, and include operators up to dimension $D\geq4$ in the Lagrangian. The couplings that appear in \re{KandU} can then be written as follows:
\bea \label{series}
  f(\varphi) &=& \sum_{n=0}^{D-2} f_n \varphi^{n} \ , \quad b_i^{(k)}(\varphi) = \sum_{n=0}^{D-4-4k} b_{in}^{(k)} \varphi^{n} \ , \quad c_1^{(k)}(\varphi) = \sum_{n=0}^{D-2-4k} c_{1n}^{(k)} \varphi^{n} \el
  c_2^{(k)}(\varphi) &=& \sum_{n=0}^{D-4-4k} c_{2n}^{(k)} \varphi^{n} \ , \quad d_1(\varphi) = \sum_{n=0}^{D-6} d_{1n} \varphi^{n} \ , \quad d_i(\varphi) = \sum_{n=0}^{D-8} d_{in} \varphi^{n} \ \text{for } i > 1 \ .
\eea

In particular, $\varphi$ could be the radial mode of the Standard Model Higgs, in which case only even powers of the field appear. We include only even powers. We first consider the case when only terms up to $D=4$ are included. Then the original potential is $V=\ha m^2\varphi^2 + \frac{1}{4} \varphi^4$, and the only non-zero coupling functions are
\bea
  f &=& M^2 + \xi \varphi^2 \ , \quad b_1^{(0)} = \text{constant} \ , \quad c_1^{(0)} \propto \varphi^2 \ , \quad c_2^{(0)} = \text{constant} \ ,
\eea
where $M$ and $\xi$ are constants. The new additive contribution to the potential is $\Delta V\propto c_1^{(0)}{}^2 \propto \varphi^4$, so it has same highest power of $\varphi$ as the original potential $V$. There is no additive contribution to the kinetic term. So there is no change to the scalar Lagrangian coming from the gauge sector, apart from a redefinition of the quartic coupling of the scalar field.

If we include operators up to dimension $D>4$, the situation changes. The potential and the kinetic term become rational functions, and it may be possible to realise different inflationary scenarios by tuning the coefficients, as in \cite{Rasanen:2018b, Langvik:2020}. Let us consider only the leading terms in \re{series} in the limit of large $\varphi$. We first look only at the terms coming from the gauge sector, neglecting the non-minimal coupling to gravity, \ie we put $f=1$. The leading behaviour of the additive contribution to the potential is $\Delta V\propto\varphi^D$, same as the original potential $V$. All of the leading terms in the numerator (and separately in the denominator) of the additive contribution $\Delta K$ to the kinetic term have the same power of $\varphi$, and we get $K\propto\varphi^{D-4}$. Defining the scalar field $\chi$ with a canonical kinetic term as
\bea \label{chi}
  \chi &=& \pm \int\rmd\varphi \sqrt{K(\varphi)} \ ,
\eea
we get the leading behaviour $\chi\propto\varphi^{\frac{D-2}{2}}$, \ie $\varphi\propto\chi^{\frac{2}{D-2}}$. In terms of the canonical field, the potential is thus $U=V+\Delta V\propto\chi^{\frac{2 D}{D-2}}$. For $D=4$ we recover the earlier result $U\propto\chi^4$. For $D=6$ we get $U\propto \chi^3$. In the limit of large $D$, when we include more and more non-renormalisable terms in the spirit of effective field theory, the potential approaches the form $U\propto\chi^2$. So higher powers of the field, due to for example quantum corrections, do not spoil the flatness of the potential, but instead make it closer to quadratic. Such a potential is by itself not viable for inflation, because even though the spectral index agrees with observations, the tensor-to-scalar ratio is above the current observational upper bound \cite{Akrami:2018odb, BICEP:2021xfz}. Including an $R^2$ term in the action will bring the tensor-to-scalar ratio down without affecting the spectral index \cite{Enckell:2018hmo}. An $R_{(\a\b)} R^{(\a\b)}$ term has the same effect, while terms higher order in $R_{(\a\b)}$ also change the spectral index \cite{Annala_thesis, Annala:2021zdt}.

Let us now look at the interplay of the non-minimal couplings in the gauge and the gravity sector. The leading contribution from the gauge sector is $\Delta K\propto\varphi^{D-4}$, which for $D>4$ dominates over the original kinetic term that is simply unity. Because of the gravity sector, the kinetic term is divided by $f\propto\varphi^{D-2}$, so we overall have $K\propto\varphi^{-2}$. Independent of the value of $D$, according to \re{chi} this leads to $\chi\propto\ln(\varphi/\varphi_0)$, where $\varphi_0$ is a constant. The potential $V+\Delta V\propto\varphi^D$ is divided by $f^2$, so we get $U\propto\varphi^{4-D}\propto e^{-(D-4)\gamma\chi}$. For $D>4$, such an exponential potential does not give slow roll in agreement with observations, unless the coefficient $\gamma$ (which is a combination of the coefficients that appear in the kinetic term) is tuned to be very small. This is a known problem for Higgs inflation if terms with ever higher powers of the field are added to the action in the Jordan frame. One proposed solution is to assume that the classical asymptotic shift symmetry in the Einstein frame extends to the quantum-corrected potential \cite{Bezrukov:2013, Rubio:2018ogq, Bezrukov:2010jz, George:2013iia, Bezrukov:2014ipa, Rubio:2015zia, George:2015nza, Fumagalli:2016lls, Bezrukov:2017dyv}.

The origin of the problem is that the non-minimal coupling $f\propto\varphi^{D-2}$ to the Ricci scalar appears quadratically in the potential, which goes like $V\propto \varphi^D$, so $V/f^2\propto\varphi^{4-D}$ asymptotes to a constant only if $D=4$. Solving the problem requires a new contribution to the potential that is quadratic in $\varphi^{D-2}$, like $f$. The non-minimal coupling to $F^\a{}_\a$ does this: we see from \re{actionE} and \re{series} that the additive contribution to the potential is proportional to $c_1^{(0)}{}^2\propto\varphi^{2(D-2)}$. However, this term is divided by the non-minimal couplings to $F_{\a\b} F^{\a\b}$ and $(F^\a{}_\a)^2$, which are $\propto\varphi^{D-4}$. If there were a principle to forbid the scalar field to couple to terms non-linear in the field strength, the potential would remain asymptotically flat for all $D$.

As inflation happens at finite field value, the contribution of higher order terms can be small, and the potential flat, if their coefficients are small. It is a quantitative question how large the coefficients can be without spoiling successful inflation. If we have non-minimal coupling only in the gravity sector, Higgs inflation is much more sensitive to higher order terms in the Palatini formulation than in the metric formulation \cite{Jinno:2019und}. This is related to the fact that although the potential is multiplied by $1/f^2$ in both cases, the kinetic term is different. In both formulations it is multiplied by $1/f$, but in the metric case there is also an additive contribution $\propto(f_{,\varphi}/f)^2$. For $f\propto\varphi^{D-2}$, this contribution is proportional to $\varphi^{-2}$ for all values of $D$. With $D>4$ this term dominates over the $1/f\propto\varphi^{2-D}$ term, and we get $\chi\propto\ln(\varphi/\varphi_0)$ for all $D\geq4$. In contrast, in the Palatini formulation, we have $\chi\propto\varphi^{\frac{4-D}{2}}$ for $D>4$. The coefficients are also different, so the formulations do not agree even for $D=4$.  As discussed above, including non-minimal coupling both in the gauge and in the gravity sector makes the field relation logarithmic for all $D\geq4$ also in the Palatini formulation, which may affect the sensitivity to higher order corrections found in \cite{Jinno:2019und}. However, this cannot be determined by considering the large field limit, but requires a more detailed analysis. 

%\vspace{0.5cm}

\section{Discussion} \label{sec:disc}

So far we have considered an Abelian gauge theory. If we instead consider a non-Abelian group, the gauge invariant term for the gauge field is $A_{\a\b}^a\equiv\pat_{[\a} A^a_{\b]}+\ha ig[A_\a,A_\b]^a$, where $a$ is a gauge group index and $g$ is the gauge coupling. The corresponding field strength $F^a_{\a\b}$ has three indices, like the affine connection $\Gamma^\c_{\a\b}$. In the gravity case all indices refer to spacetime structure, so it is possible to source the connection with any kind of field, because derivatives can provide indices. In contrast, in the non-Abelian gauge case, we need a source term with the right gauge structure to give the index $a$. If the scalar field is a gauge singlet, the only source term is $A_{\a\b}^a$, and hence $F_{\a\b}^a$ will be proportional to $A_{\a\b}^a$. Therefore there is no change in the pure scalar sector relevant for inflation. If the scalar field is in the fundamental representation (like the Standard Model Higgs), it can act as a source for $F_{\a\b}^a$ provided it appears in even powers. In this sense, there is not much difference with the Abelian case. If the scalar field and the gauge field (whether Abelian or non-Abelian) are Standard Model fields, the couplings between them in \re{actionE} are constrained by collider observations. However, the constraints are extremely weak, because the derivative couplings are suppressed by the Planck scale. For example, consider $\b_1, \b_2 \propto \varphi^2$ and $\b_3, \b_4 = $ constant. The $\b_1, \b_2$ terms lead to interactions proportional to (dropping all indices) $(E/\Mpl)^2 A^2 \varphi^2$, where $E$ is the energy, the field $A$ is either the photon, $W$ or $Z$, and we have restored the reduced Planck mass. This four-vertex is suppressed relative to the Standard model contribution by the factor $(E/\Mpl)^2\lesssim10^{-29}$ at currently accessible collider energies. The $\b_3, \b_4$ terms lead to similar interactions with one extra suppression factor of $(E/\Mpl)^2$.

Including a fermion field $\psi$ leads to a number of possible new terms. The simplest linear couplings to $F_{\a\b}$ (to generate a source term) are $\bar\psi\psi F^\a{}_\a$ and $\bar\psi\gamma^A\gamma^B\psi e^\a{}_A e^\b{}_B F_{[\a\b]}$, where $\gamma^A$ is a gamma matrix and $e^\a{}_A$ is a tetrad. Solving for the field strength, these lead to the extra contributions $F_{(\a\b)}\propto\bar\psi\psi g_{\a\b}$ and $F_{[\a\b]}\propto\bar\psi[\gamma^A,\gamma^B]\psi e_{\a A} e_{\b B}$ in the solution \re{Fsol} for $F_{\a\b}$. The symmetric contribution, when inserted back into the action, leads (at zeroth order in $X_{\a\b}$) to $\bar\psi\psi$ and $(\bar\psi\psi)^2$ multiplied by functions of the scalar field. The antisymmetric contribution leads to $\bar\psi\gamma^A\gamma^B\psi e^\a{}_A e^\b{}_B A_{\a\b}$, $\bar\psi\gamma^A\gamma^B\psi e^\a{}_A e^\b{}_B {}^*A_{\a\b}$, and $\bar\psi[\gamma^A,\gamma^B]\psi \bar\psi[\gamma_A,\gamma_B]\psi$. In the non-Abelian case, depending on the representation structure of the fermions, symmetry may prevent bare mass terms like $\bar\psi\psi$ (as happens in the Standard Model), but the term $\bar\psi\psi F^\a{}_\a$ remains allowed, and will generate couplings between the scalar field and the fermions that give mass terms, providing a mechanism to generate masses with the gauge field strength without a scalar field. Unless there is a fermion condensate during inflation, the fermion terms are not expected to be relevant during inflation, but could be important for dark matter production, as in the case of non-minimal couplings to fermions in the gravity sector \cite{Shaposhnikov:2020gts, Shaposhnikov:2020aen}.

Finally, we have assumed that the action does not contain terms higher than quadratic in the field strength $F_{\a\b}$. Let us consider what happens if we have terms of arbitrary power $F_{\a\b}$, but still only include only operators up to dimension $D$. The higher order terms lead to a non-linear equation of motion for the gauge field. After solving for the field strength the action will be non-polynomial in the gauge field, and will contain an infinite number of powers of it. For electromagnetism such terms are strongly constrained by observations. The constraints are satisfied if the coefficients of the terms vanish (or are strongly suppressed) in the electroweak vacuum, but the terms could still play a role in magnetogenesis or during preheating. During inflation, if we restrict to coefficients that are polynomial in $\varphi$, then at large fields we still have (to zeroth order in $X_{\a\b}$) $F_{(\a\b)}\propto\varphi^2 g_{\a\b}$, leading to an additive contribution to the potential that is $\propto\varphi^D$ as in the quadratic case. Similarly, the leading contribution to $K(\varphi)$ is still $\propto\varphi^{D-4}$, so there is no qualitative change to the scalar sector.

\section{Conclusions} \label{sec:conc}

We have considered a formulation of gauge field theory where the field strength $F_{\a\b}$ is an independent variable, whose symmetric part can be non-zero. We take the action to be linear in the gauge field, and include non-minimal couplings between $F_{\a\b}$ and a scalar field, considering terms linear and quadratic in $F_{\a\b}$, and linear in the kinetic term of the scalar field. As in the Palatini formulation for gravity, the field equation of the field strength then leads to a dependence on the scalar field, not only on the gauge field. The theory is physically distinct from the case where the field strength is a priori fixed in terms of the gauge field. If the scalar field is constant, we recover usual gauge field theory plus vacuum energy. Otherwise, there are new terms that involve the gauge field and the scalar field.

We consider slow-roll inflation, with the gauge field set to zero. The effect of the field strength, shifted to the scalar sector, gives an additive contribution to the kinetic term and the potential of the scalar field. If the non-minimal couplings are polynomial in the scalar field and we include only operators up to dimension $D$ in the Lagrangian, then the resulting potential is $\propto\chi^{\frac{2 D}{D-2}}$, where $\chi$ is the canonical scalar field. For $D=4$, there is no change, for $D=6$ the potential changes from the naive result $\chi^6$ to $\chi^3$, and for higher powers of $D$ the potential approaches $\chi^2$. Adding higher order powers from loop corrections thus does not spoil the flatness of the potential, but instead makes it closer to quadratic. A quadratic potential leads to a tensor-scalar-ratio above the observational upper limit, which can be lowered by including a $R^2$ term in the action if gravity is also described in the Palatini formulation \cite{Enckell:2018hmo}.

In Higgs inflation, which relies on non-minimal coupling between gravity and the Higgs field, the potential is asymptotically flat if only operators up to $D=4$ are included. In the Palatini formulation for gravity, Higgs inflation is more sensitive to higher order terms that break the asymptotic flatness than in the metric formulation \cite{Jinno:2019und}. The reason is that the mapping between the Jordan frame Higgs field and the Einstein frame canonical field is different. In the large field limit, in the Palatini case the relation between the Jordan frame Higgs field and the canonical field is a power-law unless $D=4$ when it is logarithmic, whereas in the metric case it is logarithmic for all $D\geq4$, and the coefficients are also different. With the non-minimal gauge field couplings we have discussed, the relation is logarithmic in the Palatini case as well for all $D\geq4$. It would be interesting to study whether this has an impact on the sensitivity to higher order corrections. If the scalar field couples only to terms linear in $F_{\a\b}$, the asymptotic flatness of the potential is preserved exactly both in the metric and in the Palatini formulation of gravity.

If terms higher than quadratic in $F_{\a\b}$ are included in the action (keeping the couplings polynomial), the situation remains qualitatively the same for the scalar field. However, the gauge field part of the action becomes non-polynomial, but may still be well-behaved. If derivatives of the field strength appear in the action, we may have extra degrees of freedom as in the gravitational Palatini case, and the resulting theory can be unstable. The modifications to the gauge field sector could be important for magnetogenesis during inflation. They could also affect the duration of reheating and thus have an impact on inflationary predictions.

\newpage

\section*{Acknowledgments}

\noindent S.R. thanks Matti Heikinheimo for a useful discussion. Y. V. gratefully acknowledges the Helsinki Institute of Physics of the University of Helsinki for hospitality during the visit where this work was carried out.

\appendix \label{app:ab}

The coefficients in the solution \re{Fsol} for $F_{[\a\b]}$ are 
\bea
  \a_1 &=& - \frac{a_1 b_2 - a_2 b_3}{2 ( b_2^2 - b_3^2 )} \el
  && + \left[ \frac{ (a_1 b_2 - a_2 b_3) b_3 [ b_2 (d_6 - d_7) + b_3 (-d_3 + d_4 + d_5) ] }{ 4 b_2( b_2^2 - b_3^2 )^2 } - \frac{ 2 a_2 b_3 (-d_3 + d_4 + d_5) + a_2 b_2 (d_6 - d_7) + 2 (a_5 - a_6) b_2 b_3 }{ 8 b_2 ( b_2^2 - b_3^2 ) } \right] X \el 
  \a_2 &=& \frac{a_1 b_3 - a_2 b_2}{2 ( b_2^2 - b_3^2 )} \el
  && + \left[ - \frac{ ( a_1 b_3 - a_2 b_2 ) b_3 [ b_2 (d_6 - d_7) + b_3 (-d_3 + d_4 + d_5) ] }{ 4 b_2 ( b_2^2 - b_3^2 )^2 } + \frac{ b_3 [ a_2 (-d_6 + d_7) + 2 ( a_5 - a_6 ) b_3 ] }{ 8 b_2 ( b_2^2 - b_3^2 ) } \right] X \el
  \a_3 &=& - \frac{ a_1 ( - d_3 + d_4 + d_5 ) + ( a_3 - a_4 ) b_2 + ( a_5 - a_6 ) b_3 }{2 (b_2^2 - b_3^2)} \el
  \a_4 &=& \frac{ ( a_1 b_2 - a_2 b_3 ) ( d_6 - d_7 ) + 2 ( a_1 b_3 - a_2 b_2 ) ( -d_3 + d_4 + d_5 ) - 2  ( a_5 - a_6 )( b_2^2 - b_3^2 ) }{ 4 b_2 (b_2^2 - b_3^2) } \el
  \a_5 &=& - \frac{ ( a_1 b_2 - a_2 b_3 ) (d_6-d_7) + 2 a_1 b_3 ( - d_3 + d_4 + d_5 ) + 2  ( a_3 - a_4 ) b_2 b_3 + 2 ( a_5 - a_6 ) b_3^2 }{ 4 b_2 (b_2^2 - b_3^2) } \ .
\eea
Note that the coefficients $a_i$ and $b_i$ contain parts proportional to $X$.

The coefficients of the gauge field terms in the action \re{actionE} are
\bea
  \b_1 &=& a_1 \a_1 + a_2 \a_2 + b_2 ( \a_1^2 + \a_2^2 ) + 2 b_3 \a_1 \a_2 \el
  && + \ha \left[ - a_2 \a_4 + ( - a_5 + a_6 ) \a_2 - 2 b_2 \a_2 \a_4 - 2 b_3 \a_1 \a_4 + ( - d_3 + d_4 + d_5 ) \a_ 2^2 + ( - d_6 + d_7 ) \a_1 \a_2 \right] X \el
  \b_2 &=& a_1 \a_2 + a_2 \a_1 + 2 b_2 \a_1 \a_2 + b_3 ( \a_1^2 + \a_2^2 ) \el
  \b_3 &=& - a_1 \a_3 + a_2  ( \a_4 + \a_5 ) + ( - a_3 + a_4 ) \a_1 + ( a_5 - a_6 ) \a_2 + 2 b_2 ( - \a_1 \a_3 + \a_2 \a_4 + \a_2 \a_5 ) \el
  && + 2 b_3 ( \a_1 \a_4 + \a_1 \a_5 - \a_2 \a_3 ) + ( - d_3 + d_4 + d_5 ) ( \a_1^2 - \a_2^2 ) \el
  \b_4 &=& a_1 ( - \a_4 + \a_5 ) - a_2 \a_3 + ( - a_3 + a_4 ) \a_2 + ( - a_5 + a_6 ) \a_1 + 2 b_2 ( - \a_1 \a_4 + \a_1 \a_5 - \a_2 \a_3 ) \el
  && +  2 b_3 ( - \a_1 \a_3 - \a_2 \a_4 + \a_2 \a_5 ) + 2 ( - d_3 + d_4 + d_5 )  \a_1 \a_2 + ( - d_6 + d_7 )  ( \a_1^2 + \a_2^2 ) \ .
\eea
Note that the coefficients $a_i$, $b_i$, and $\a_i$ contain parts proportional to $X$.

\bibliographystyle{JHEP}
\bibliography{pem}

\end{document}